# H induced decohesion of an Al grain boundary investigated with first principles: Proposed general conditions for instant breakage and delayed fracture

F. J. H. Ehlers[1,2,*], M. Seydou[1], D. Tingaud[2], F. Maurel[1], Y. Charles[2], S. Queyreau[2]

[1]University Paris Diderot, Sorbonne Paris Cité, ITODYS, UMR 7086 CNRS, 15 rue J.-A. de Baïf, 75205 Paris cedex 13, France

[2]Université Paris 13, Sorbonne Paris Cité, Laboratoire des Sciences des Procédés et des Matériaux, LSPM, CNRS, UPR 3407, 99 avenue Jean-Baptiste Clément, F-93430 Villetaneuse, France

**(Abstract)**

The uniaxial tensile test response of a H decorated $\Sigma 5$ $[100]$ twist grain boundary (GB) in face-centered-cubic Al has been examined with first principles. The impurity shows a strong tendency to relocate during loading. To capture these H movements, the standard model framework was extended to probe loading-unloading hysteresis. If the maximum tensile stress accepted by the H decorated GB in the slow fracture limit is reached before the maximum acceptable strain, exceeding this stress may trigger a H influx-controlled destabilization, as opposed to 'immediate' breakage. Such 'delayed' failure appears likely whenever the H attraction to a GB displays a monotonic decrease with increased loading.

**(Bulk text)**

*Hydrogen induced decohesion* (HID) is a prominent, yet controversial [1] member among the candidates for hydrogen embrittlement (HE). When HID is operative, H impurities agglomerate at locations of high hydrostatic stress, drastically reducing the strength of adjacent metal host atom bonds, thus in turn affecting the material toughness and ductility – the trademarks of HE. Atomistic simulations generally support a detrimental effect of H; see, e.g., [2]. Experimentally, an increased crack tip sharpness with H concentration has been observed in cleavage studies for some materials; see, e.g., [3]. However, the *relative importance* of HID remains debated, with complications arising from two factors: (i) HE in general incorporates both plastic and elastic contributions. Consequently, HID is in direct competition with other mechanisms [1]. (ii) Detection of H *in situ* is highly challenging [4]. This fact obstructs an experimental analysis where the impurity influence during a fracture process is quantified with reference to local concentration. Technically, theoretical studies are not limited in the same manner and hence may provide the missing pieces of the puzzle if applied in conjunction with experiment. In the case of a HE process dominated by HID, the crack tip would tend to respond purely elastically. Here, *ab initio* modelling applied to the heart of the fracture region may generate accurate results if appropriately coupled structurally and chemically to the surrounding grains. This consideration would appear particularly intriguing for the scenario of intergranular fracture, where a satisfactory determination of the H decorated grain boundary (GB) loading response is potentially out of reach for semi-empirical potentials. In the absence of detailed knowledge on the H-GB interactions during loading, large-scale schemes (e.g., using finite element modelling) have so far resorted to treating the GB as a simple 2D surface [5]. The present paper aims at analyzing the general H influence on a decohering metal GB at the atomistic level, with basis in first principles simulations. Two key points are revealed: (i) Trapped H may prefer to change sites during loading, rendering single-site models inadequate. (ii) If this H trapping is persistently strengthened as breakage is approached, the final stage before fracture may be H diffusion limited.

The thermodynamic foundations to a quantitative understanding of HID were laid roughly four decades ago, with the establishment of process rate dependent equations for the general problem of an impurity decorated GB subjected to uniaxial tensile stress [6, 7]. In the limits of both slow and fast fracture, determination of the fracture energy required for separating the two grains was shown to require information from the extreme points of the fracture process only. In other words, the sole GB related contribution to this quantity would involve the zero-load configurations. This conclusion reflects the loading response of the conjugate variables ($\mu$, $\Gamma$) – the impurity chemical potential and surface density, respectively. In the fast fracture limit, any impurity influx to the GB is effectively suppressed (fixed $\Gamma$), while at the other extreme, the entire system remains in thermodynamic equilibrium (fixed $\mu$). Generally, these parameters should be determined with reference to the *GB vicinity* – the GB region of influence from the perspective of the impurity. $\Gamma$ is defined as the number of impurity atoms within this volume, divided by the GB area, while $\mu$ is the chemical potential value immediately outside the vicinity. Due to the high mobility of H at ambient conditions, it is questionable if either of the above limits can be applied to this case [6]. When impurity mobility and crack speed are comparable, both $\mu$ and $\Gamma$ would be expected to vary during the fracture process [7], reflecting the system attempt to preserve chemical equilibrium when the GB displays an evolving ability to attract impurities. Here, an efficient modelling requires both a multi-scale scheme [8–10] and a careful atomistic examination of the impurity decorated GB structure over the full loading range to fracture.

Until now, *ab initio* simulations of H decorated metal GBs subjected to loading [2, 11, 12, 13] have addressed the fast fracture limit only. Some of these studies, concerning body-centered-cubic Fe, have examined the H influence on the GB energy evolution with loading – providing information linked to the GB cohesive stress $\sigma$ in addition to the fracture energy. The predicted changes to $\sigma$ were weak, indicating that the chosen GB and material may not be optimal for highlighting HID effects related to intact GBs. The case of face-centered-cubic (fcc) Al appears different, as potentially supported by experiments [14]. Recently, a highly favorable H site was predicted [15] upon subgrain displacement in bulk Al. The impurity was located at a bridge (B) site between a metal host pair, with a coordination number of two. For most GBs, such a site should appear upon sufficient loading, as a consequence of sequential bond breakage and increasing alignment of remaining bonds with the GB normal. In this work, an *ab initio* uniaxial tensile test has been performed for a fcc Al $\Sigma 5$ [100] 36.87° twist GB (TGB) in the presence of H. The selected system has been described in detail in [16], where cell size convergence and computational precision were carefully addressed. For the sake of simplicity, these conclusions were presumed unaffected by H presence. The chosen GB supercell hosted 40 Al atoms and two GBs. One H atom was deposited at each GB, producing $\Gamma = 2.5 \times 10^{-2}$ Å$^{-2}$ (labeled $\Gamma_0$ below). At zero load, the impurity-free system was fully relaxed, and the H influence on the cell dimensions was neglected throughout. The resulting simulation accuracy is considered satisfactory for the intended discussion of trends. The density functional theory calculations have used the projector augmented wave method for the description of the electron-ion interactions [17], as implemented in the Vienna *Ab initio* Simulation Package (VASP) [18]. The exchange-correlation functional was described in the Perdew-Burke-Ernzerhof generalized gradient approximation [19].

From the atomistic perspective, full information on HID in a uniaxial tensile test may be obtained by assessing the H influence on the *GB traction-separation (T-S) curve*, i.e., the variation in $\sigma$ with the separation (elongation) of the grains $\delta$. In addition to the fracture energy, the T-S curve highlights the entire stress-strain evolution, especially the maximum ('critical') stress $\sigma_c$ that the GB can withstand. In practice, the T-S curve construction involves the quantification of two linked aspects: The attraction of H to the GB, and the expected values of $\Gamma$ during a given loading procedure. The latter question, while as central as the former, is connected with multi-scale modelling and hence will only be qualitatively touched upon here. For the isolated GB supercell study, two scenarios are straightforwardly addressed [7]. In the limit of slow fracture, at fixed temperature $T$, the GB cohesive stress is obtained as

$$\sigma = (1/A)(\partial \Omega / \partial \delta)_{T,\mu}, \tag{1}$$

while in the limit of fast fracture

$$\sigma = (1/A)(\partial G / \partial \delta)_{T,\Gamma}. \tag{2}$$

In these equations, $A$ is the GB area while $\Omega$ denotes the grand canonical potential for the GB vicinity, related to the Gibbs free energy $G$ by

$$\Omega = G - \mu A \Gamma. \tag{3}$$

Generally, *ab initio* simulations neglect the conjugate variable terms $-TS$ and $PV$, replacing $G$ by the internal energy $E$. This work adopts the same strategy. A full incorporation of the pressure-volume ($PV$) term would require that a self-consistent structural coupling to the stress field in the surrounding grains be made [7]. For the temperature-entropy ($TS$) term, configurational entropy changes (expected weak) could be formally included in zero-temperature studies. Eq. (1) and (2) are shown below to play a central role also when discussing fracture processes beyond the slow and fast limits.

The level of H segregation at the GB is clarified via the evolution with $\delta$ in the impurity formation energy, $E_H^f(i)$, i.e., the energy required to take the H atom from its presumed initial state as part of an environmental H₂ molecule and onto a site $i$ at the GB. Formally associating the H surface density $\Gamma_0$ with the dilute limit (isolated H),

$$E_H^f(i, \Gamma_0, \delta) = E(i, \Gamma_0, \delta) - E(\delta) - \frac{1}{2} E(H_2) - \mu^*. \tag{4}$$

In Eq. (4), the first two energies on the right-hand side are per GB, i.e., half a supercell energy. Only the former cell incorporates H. The chemical potential value $\mu^*$ is computed relative to $E(H_2)$ – the energy of an isolated H₂ molecule. As shown below, $\Gamma_0$ was found to be high compared to realistic conditions. Consequently, no direct modelling of H-H interactions was performed. Theoretical surface densities $\Gamma_i$ in the slow fracture limit were computed using the Langmuir-McLean equation [20], as implemented in [21]. Using Eq. (3) and (4), the grand canonical potential reads

$$\Omega(\Gamma, \delta) = E(\delta) + \sum_i \left[ (\Gamma_i / \Gamma_0) E_H^f(i, \Gamma_0, \delta) \right], \tag{5}$$

with

$$\Gamma = \sum_i \Gamma_i. \tag{5a}$$

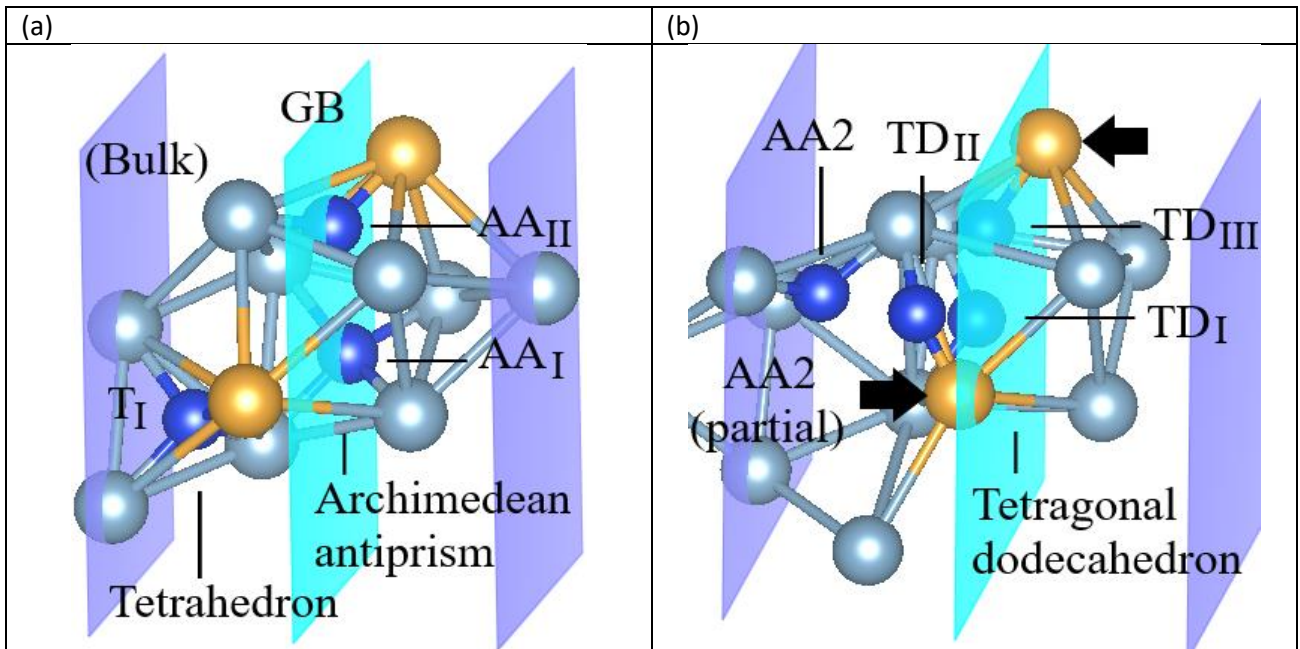

FIG. 1. Computed tensile stress response of the fcc Al $\Sigma 5$ [100] TGB structure (key Al movements highlighted with arrows and colors), and effect on the set of competitive H sites. (a): $\delta$ =0 Å. (b): $\delta$ =1.5 Å. H atom (dark blue spheres) positions are approximate, with impurity induced distortions suppressed throughout.

When performing an *ab initio* uniaxial tensile test on a GB free of impurities, it is recommended that the point of breakage be approached from the state of zero load via a sequential increase in $\delta$ [8]. In this 'dragged elongation' scheme (DES), the atomic positions should be optimized for each chosen $\delta$. For a GB decorated with H, adopting the same strategy may produce incorrect answers, however. This is exemplified by the Al $\Sigma 5$ [100] TGB, which undergoes a structural transformation during the loading stage, linked to the 'bump' on the T-S curve in [22]. This transition implies alterations (Fig. 1) to the GB characteristic set of polyhedral units [23] and, quite possibly, to the set of competitive H sites. With the absence of dynamics in the DES, there is no guarantee that this scheme will reveal every relevant site generated. H relocation to a site clearly favored at high loading may be prevented by the persistent presence of even weak energy barriers at each $\delta$. Given this shortcoming, the DES was accompanied by a study tailored to identify all sites at GB separations *above* the structural transformation ($\delta \geq 1.35$ Å). Starting from an identification of all competitive H sites at $\delta$ =1.5 Å, the new scheme includes both a DES and a 'gradual compression' scheme (GCS), reverse in nature to the DES. The level of hysteresis (energy barrier influence) in the system can now be probed as the difference between the two scheme results, as the GCS and original DES have the same basic limitation.

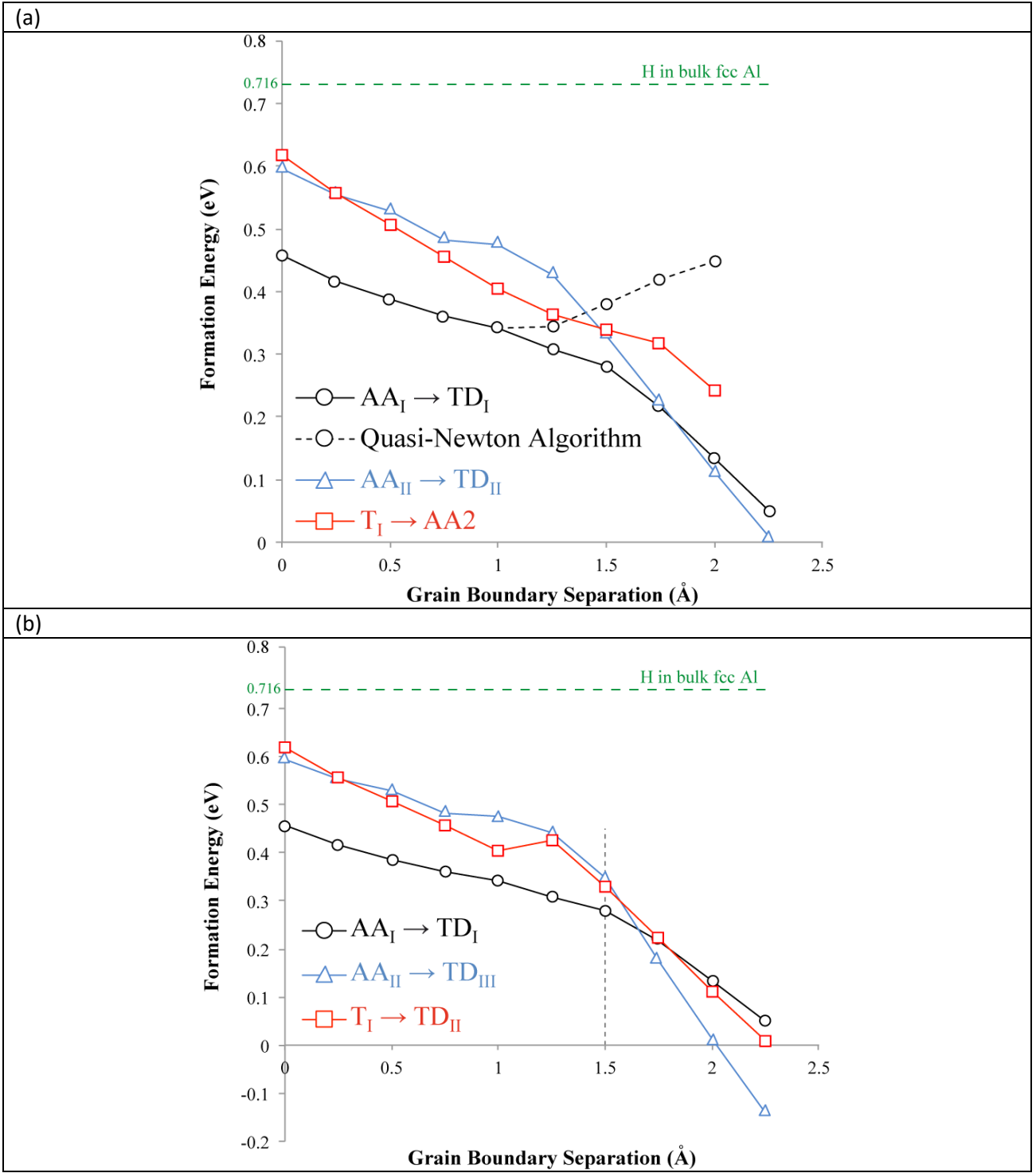

FIG. 2. Calculated H formation energies ($\mu^*$ =0 eV) at the fcc Al $\Sigma 5$ [100] TGB structure, obtained using (a) the DES and (b) the DES-GCS (centered on the vertical dashed line). Quantitative differences are noted.

Figure 2(a), 2(b) show the computed $E_H^f(i, \Gamma_0, \delta)$ obtained using the original DES and the new DES-GCS, respectively. Each study has focused only on the sites within 0.2 eV of the most favorable formation energy at the respective scheme starting points [24]. Both schemes were applied until $\delta$ =2.25 Å

– above the impurity-free GB 'ideal' breakage point [22]. The preferred H sites remain in the same geometry throughout (see Fig. 1), but the geometry itself is changing, from an Archimedean antiprism (AA) in Fig. 1(a) to a tetragonal dodecahedron (TD) in Fig. 1(b). The impurity atoms relocate in response, occupying interstitial sites ($AA_I$, $AA_{II}$) at low loading, but moving to B sites ($TD_I$, $TD_{II}$, and $TD_{III}$) as breakage is approached, in accordance with the expectations from [15]. Only the DES-GCS reveals the most favorable site at high loading, $TD_{III}$. The relative stability of the different sites is strongly affected by the tensile stress, indicating a general need for examining more than one site. These results emphasize the expected general usefulness of a combined DES and DES-GCS analysis for H decorated metal GBs subjected to loading [25]. Hysteresis is prominent in Fig. 2, with, e.g., the DES predicting a H movement from the best tetragonal site $T_I$ to a site in a *different* AA geometry, AA2 (see Fig. 1(b)). Dynamical simulations, beyond the scope of this work, would be required to quantify the influence of such metastable configurations, including the ease of transition to the $TD_{III}$ site.

For the T-S curve calculations, the thermodynamics-based analysis of this work requires that all transition barriers in Fig. 2 be neglected, producing well-defined $E_H^f(i, \Gamma_0, \delta)$ values as input for the H surface density determination. H in Al is essentially an ideal solution for all scenarios of expected relevance, with the relation between H bulk concentration $c_H$ and $\mu^*$ outlined in [26]. For realistic $c_H$ values and $T$ =300 K, the GB remains H-free at zero load, consistent with experimental findings [14]. This result renders the fast fracture limit trivial, with only the slow fracture limit discussed below.

The T-S curve emerging from Eq. (1) has the inherent limitation [8] that $\Omega$ has $\delta$ as a controlling variable, whereas this parameter is more suitably described as the response to a well-defined external stress $\sigma$. The grand-force potential $\Phi$ appropriate for the latter consideration is related to $\Omega$ as

$$\Phi = \Omega - \delta \left( \partial \Omega / \partial \delta \right)_{T,\mu}. \tag{6}$$

This connection implies that certain states on the curve described by Eq. (1) may be purely hypothetical. If two stable configurations on this curve are linked by the same cohesive stress, and if the work required to move between these configurations at fixed stress is identical to the value obtained via Eq. (1), Eq. (6) predicts a H influx assisted transition at constant stress $\sigma_t$. The GB is still able to withstand a higher external stress, as the value of $\delta$ where the transition is completed marks the onset of a rising $\sigma$ in general. In other words, subjecting the GB vicinity to an external stress $\sigma_t$ does not trigger breakage. Figure 3 shows an example for the impurity-free Al $\Sigma 5$ [100] TGB.

A local stress maximum occurring before the GB breakage point in Eq. (1) however may also be real. If this maximum is visited, it represents the critical stress $\sigma_c = \sigma_c(\Gamma_c)$ of the GB in question, since a $\sigma_t$ can always be found otherwise. Consequently, the GB is destined to break when reaching the GB separation $\delta_c$ (Fig. 3). The exact mechanism depends on the system evolution at $\delta > \delta_c$. As the external stress is raised slightly above $\sigma_c$, the GB enters a stage of mechanical instability, i.e., a *fast* transformation during which H influx and external stress changes can be neglected. If mechanical equilibrium is reestablished at some higher value of $\delta$ for this system, breakage remains ensured by the GB vicinity no longer being in *chemical equilibrium* with the surrounding grains, but depleted of H: Any subsequent H influx will destabilize the GB according to Eq. (1), for which $\sigma(\delta) < \sigma_c$ for $\delta > \delta_c$. In other words, H does not necessarily induce 'immediate' breakage at $\sigma = \sigma_c$ in this scenario. The alternative 'H diffusion controlled (HDC) delayed fracture' mechanism would seem both likely and influential if the strong monotonic decrease in $E_H^f(i)$ in Fig. 2 represents a typical H-GB interaction behavior. The GB cohesive stress during mechanical instability is modeled here using Eq. (2) with a 'frozen' H distribution $\Gamma_i = \Gamma_{c,i}$, i.e., the process is deemed too fast for impurity relocation. As the H sites are altered during this stage, practical analysis refers to the DES results.

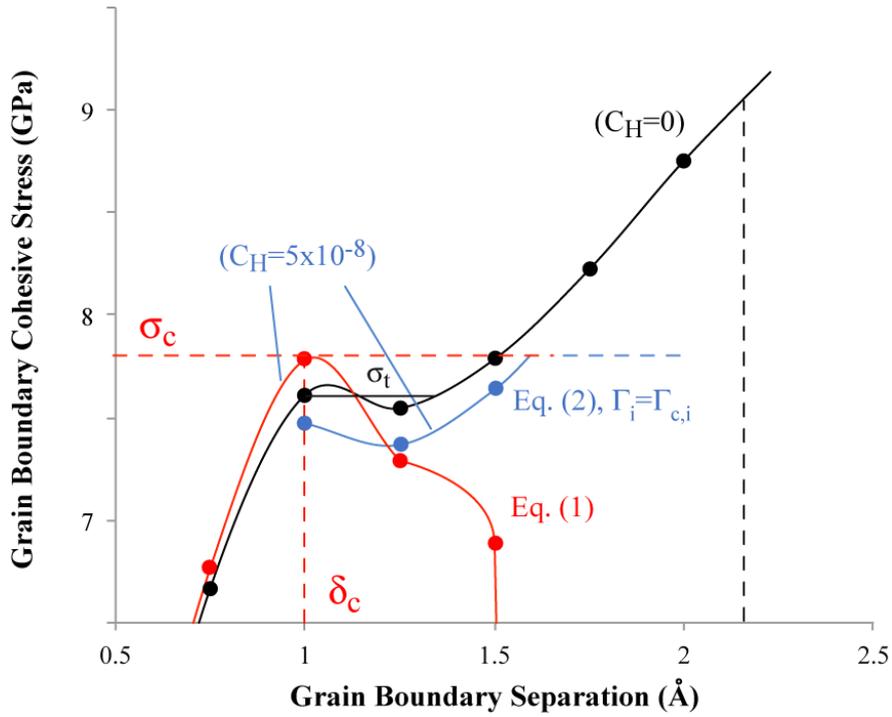

FIG. 3. Calculated H effect on the fcc Al $\Sigma 5$ [100] TGB T-S curve in the slow fracture limit, $c_H = 5 \times 10^{-8}$ (impurity-free curve also shown). Eq. (1) remains valid until $\delta = \delta_c$, while Eq. (2) must obey $\sigma \leq \sigma_c$. All curves are guides to the eye.

The calculations for the Al $\Sigma 5$ [100] TGB show that HID for the GB vicinity may be triggered in the slow fracture limit, with breakage occurring in this case by HDC delayed fracture. Figure 3 describes such a scenario, at $c_H = 5 \times 10^{-8}$ (site$^{-1}$). The calculated critical stress ($\delta_c = 1.0$ Å) is 7.78 GPa – a reduction of at least 14% from the value (at $\delta = 2.15$ Å) in the absence of H [22]. The H surface density during the mechanical instability stage is $\Gamma_c = 9.4 \times 10^{-3}$ Å$^{-2}$ ($\approx 0.4 \Gamma_0$), with 72% of the atoms on the AA$_l$ (later TD$_l$) sites. The GB cohesive stress reaches 7.78 GPa again at $\delta \approx 1.59$ Å, with $\sigma \leq \sigma_c$ expected at larger GB separations. Analytically derived H stress contributions from Eq. (1) and (5) indicate that $E_H^f(i)$ can be at most a few tens of meV for H located at site $i$ to cause a $\sigma$ reduction. As expected, also $\partial E_H^f(i)/\partial \delta$ is influential. In the simulations, the large numerical values for this quantity induce a non-negligibly *raised* $\sigma_c$ in a range below $c_H = 2.3 \times 10^{-8}$. If the formation energies remain monotonically decreasing when simulation precision is increased, the qualitative conclusions outlined here should remain unchanged. The chosen chemical potential $\mu^*$ =0.306 eV in Fig. 3 corresponds to a rather appreciable equivalent H$_2$ pressure ($\approx 6$ GPa) [27].

An HDC delayed fracture criterion may be formulated with application to any computationally tractable metal GB subjected to a uniaxial tensile test. If all configurations up to $\delta_c$ are visited in the slow fracture limit, this fracture process is predicted for an external stress slightly above $\sigma_c$, provided that the inequality

$$(1/A)\sum_i \left[ \left( \Gamma_{c,i} / \Gamma_0 \right) \partial E_H^f(i) / \partial \delta \right] \geq \sigma_c(\Gamma_c) - \sigma(\delta) \qquad (7)$$

is satisfied for a realistic value of $\delta > \delta_c$. Here, the stress $\sigma(\delta)$ refers to the *impurity-free* GB and may exceed $\sigma_c(\Gamma_c)$. Eq. (7) may also be applied to the case where $\sigma_c$ is reached via a faster process, and with the external stress kept fixed subsequently. This may affect $\Gamma_{c,i}$ (and, consequently $\sigma_c$), but the condition is unchanged as long as no path towards thermodynamic equilibrium at $\delta > \delta_c$ is implied by Eq. (1). This extended range of application of Eq. (7) technically allows for placing conservative boundaries on critical external influences and system treatments, even before attempting full integration of the *ab initio* analysis in a multi-scale model scheme.

**(Acknowledgments)**

The calculations were performed using HPC resources from GENCI- [CCRT/CINES/IDRIS] (Grant 2016-[t2016097681]) and MAGI, the HPC cluster of SPC (Sorbonne Paris Cité) University. The authors thank N. Greneche for his support on MAGI. ANR (Agence Nationale de la Recherche) and CGI (Commissariat à l'Investissement d'Avenir) are gratefully acknowledged for their financial support through Labex SEAM (Science and Engineering for Advanced Materials and devices), ANR 11 LABX 086, ANR 11 IDEX 05 02 and through the funding of the MMEMI (Multi-scale modelling of materials and interfaces) project. Figure 1 was constructed using the VESTA software [28].